\documentclass{aa}
\usepackage{graphicx}
\usepackage{natbib}
\newcommand{\Hb}{\ensuremath{{\rm H}\beta}}

\newcommand{\Mgb}{\ensuremath{{\rm Mg}\, b}}

\newcommand{\Fe}{\ensuremath{\langle {\rm Fe}\rangle}}

\newcommand{\aFe}{\ensuremath{\alpha/{\rm Fe}}}

\newcommand{\ZH}{\ensuremath{Z/{\rm H}}}

\begin{document}

\titlerunning{\aFe\ enhanced stellar tracks and the ages of elliptical
galaxies}

\title{\boldmath The impact of \aFe\ enhanced stellar evolutionary
tracks on the ages of elliptical galaxies}

\authorrunning{D.~Thomas
         \and 
        C.~Maraston}

\author{Daniel Thomas
         \and 
        Claudia Maraston}

\offprints{Daniel Thomas}
\institute{Max-Planck-Institut f\"ur extraterrestrische Physik,
Giessenbachstra\ss e, D-85748 Garching, Germany\\
\email{dthomas@mpe.mpg.de, maraston@mpe.mpg.de}}

   \date{Received 16 September 2002; accepted 30 January 2003}

   \abstract{ We complement our study of \aFe\ enhanced stellar
population models of Lick absorption indices \citep{TMB03} by
comparing two sets of \aFe\ enhanced models. In both models the impact
on Lick indices due to \aFe\ enhancement is accounted for through a
modification of the stellar absorption line-strengths using the
response functions of \citet{TB95}.  One set of models, however, uses
solar-scaled, the other \aFe\ enhanced stellar evolutionary
tracks. Since the \aFe\ enhanced tracks are {\it hotter} than the
solar-scaled ones \citep{Saletal00}, the correspondent stellar
population models have slightly {\em weaker} metallic indices (i.e.\
\Mgb, \Fe\ etc.) and {\em stronger} Balmer line indices (\Hb)
\citep{Maretal03}.  Here we explore quantitatively the impact of this
effect on the \aFe\ ratios, metallicities and ages that are derived
for elliptical galaxies.  We find that the modest decrease of the
metallic indices \Mgb\ and \Fe\ balance each other, such that fully
consistent \aFe\ ratios are derived for stellar systems using \aFe\
enhanced models with either solar-scaled or \aFe\ enhanced stellar
tracks.  The decrease of the metallic indices and the increase of \Hb\
conspire in a way that also consistent metallicities are obtained.
The derived ages, instead, are significantly different. The inclusion
of \aFe\ enhanced stellar tracks leads to the derivation of ages as
high as $\sim 30$~Gyr for elliptical galaxies. For the same objects,
ages not older than 15~Gyr are obtained, if \aFe\ enhanced models
using solar-scaled tracks are adopted.  This may indicate that current
stellar evolutionary models overestimate the bluing of stellar
evolutionary tracks due to \aFe\ enhanced chemical mixtures at
super-solar metallicities.

   \keywords{
stars: evolution -- stars: abundances -- galaxies: stellar content --
galaxies: elliptical and lenticular, cD -- galaxies: formation
               }

  }

   \maketitle
%
%________________________________________________________________

%================= I N T R O D U C T I O N ===============================

\section{Introduction}
The analysis of elliptical galaxy absorption lines in the early 90s
revealed an inadequacy of standard models to recover the Mg and Fe
Lick indices measured in these galaxies.  The failure was attributed
to element abundance ratios different from solar. More specifically,
an enhanced ratio of $\alpha$-elements (e.g. Mg, O, Ca, etc.) with
respect to iron was the most straightforward interpretation
\citep{WFG92}. Recently an accurate model calibration with metal-rich
globular clusters has confirmed the early suggestion
\citep{Maretal03}. In fact, standard stellar population models are
found to be unable to recover the indices also of Bulge globular
clusters that are known, from high-resolution spectroscopy of
individual stars, to have enhanced \aFe\ ratios. These findings call
clearly for an update of the standard models, in particular for the
inclusion of \aFe\ enhancement.

Therefore, in \citet*[][hereafter TMB]{TMB03} we present the Lick
absorption indices of simple stellar population models (SSP) which,
for the first time, are computed for variable element abundance
ratios, in particular \aFe.  The key ingredient in the models is the
response of absorption lines to the abundance variations of individual
elements in the stellar atmospheres. This is the major effect to be
taken into account \citep[see also][]{Traetal00a}. The secondary
effect of \aFe\ enhancement on SSP models comes from the adopted
stellar evolutionary tracks, since the element abundance variations in
a star affect also the star's evolution and opacities, hence the
effective temperature.  This secondary effect has been neglected in
the TMB models, because they are based on the SSPs of \citet{Ma98},
which use solar-scaled tracks \citep{CCC97}. However, the TMB models
are calibrated with globular clusters and reproduce very well their
metallicities and \aFe\ ratios in the metallicity range up to
solar. This does not come as a surprise, since the difference between
\aFe\ enhanced and solar-scaled stellar evolutionary tracks is
negligible at low metallicities
\citep{SW98,Saletal00,Vanetal00,Kimetal02}.

A larger difference is found at super-solar metallicities
\citep{Saletal00}.  More specifically, \aFe\ enhanced stellar tracks
have hotter turnoffs and hotter giant branches owing to the lower
opacities of \aFe\ enhanced stellar atmospheres
\citep{Saletal00,Vanetal00,Kimetal02}.  \citet{Maretal03} explore the
effect on SSP models.  They show that SSPs using \aFe\ enhanced tracks
have moderately weaker metallic (e.g., \Mgb, \Fe) indices, and
significantly stronger Balmer (\Hb) indices for the same age and
metallicity.  The aim of this paper is to estimate this effect on the
TMB models, and to discuss the impact on the ages, metallicities, and
\aFe\ ratios derived for elliptical galaxies. 

The paper is organized as follows. Sect.~2 gives a brief overview of
the development and current status of \aFe\ enhanced stellar
evolutionary tracks. The modified TMB models are described in
Sect.~3 and compared with elliptical galaxy data in Sect.~4. The
results are summarized and discussed in Sect.~5.

%=============== A L P H A - E N H A N C E D  T R A C K S =============

\section{\boldmath A brief history of \aFe\ enhanced stellar tracks}
The development of stellar evolutionary tracks with chemical mixtures
different from the solar proportions has made major progress in the
last years.  Based on the assumption that the abundance ratio of
elements with high ionisation potential (HPE: C, N, O, Ne) to the
elements with low ionisation potential (LPE: Mg, Si, S, Ca, Fe)
remains constant relative to the solar ratio in an \aFe\ enhanced
isochrone, \citet{SCS93} find that an \aFe\ enhanced isochrone is
cooler than the solar-scaled one and can be mimicked by a solar-scaled
isochrone with higher metallicity.  The reason for the validity of
this metallicity-scaling is that the HPE control the temperature of
the turnoff, while the LPE impact on the red giant branch \citep[see
also discussion in][]{Traetal00a}, so that the constancy of the
HPE/LPE ratio ensures the shape of the stellar evolutionary track not
to vary.

But the ratio of HPE to LPE is actually super-solar for realistic
\aFe\ enhanced mixtures, and the shape of the isochrone is not
conserved \citep{WPM95}. Most importantly, \citet{SW98} find that for
$[{\rm HPE/LPE}]\approx 0.1$ the turnoff and also the red giant branch
of an \aFe\ enhanced isochrone are {\em bluer} than that of a
solar-scaled one, in contrast to the results of \citet{SCS93}.
However, the study of \citet{SW98} is restricted to sub-solar
metallicities.  Recently, \citet{Saletal00} published a set of \aFe\
enhanced ($[\aFe]=0.3$) stellar tracks for super-solar
metallicities. The authors confirm the conclusion found by
\citet{SW98} at low metallicities: '{\rm at the same evolutionary
stage, the solar-scaled track is fainter, cooler, and older than the
$\alpha$-enhanced one}' \citep{Saletal00}. \citet{Vanetal00} and
\citet{Kimetal02} come to the same conclusion.

\section{The modified TMB model}
We consider the synthetic Lick indices of three different sets of
SSP models (see Table~\ref{tab:models}). 
\begin{enumerate}
\item The TMB model for $[\aFe]=0.3$ (Model~1).
\item A modification of Model~1 that includes additionally the effect
of \aFe\ enhanced stellar evolutionary tracks (Model~2).
\item The standard solar-scaled ($[\aFe]=0.0$) model (Model~3).
\end{enumerate}

\begin{table}
\caption{The three model flavors}
\begin{tabular}{cccc}
\hline\hline
   &        & \multicolumn{2}{c}{Effect of \aFe\ on}\\
   & [\aFe] & Absorption Lines & Stellar Tracks \\
\hline
Model 1 & 0.3 & yes & no\\
Model 2 & 0.3 & yes & yes\\
Model 3 & 0.0 &  no & no\\
\hline
\end{tabular}
\label{tab:models}
\end{table}

Model~1 and Model~2 have the enhanced \aFe\ ratio $[\aFe]=0.3$. In
both, the element abundance effects on the strength of the absorption
features, i.e.\ on the Lick indices, are taken into account as
described in detail in TMB. The impact of \aFe\ enhancement on the
underlying stellar evolutionary tracks, instead, is neglected in
Model~1 but considered in Model~2 as explained in the following.

In both models the stellar evolutionary tracks (solar-scaled) are
taken from \citet{CCC97} and S.~Cassisi (1999, private
communication). As these authors do not yet provide stellar tracks for
non-solar element ratios, in Model~2 we mimic the effect of \aFe\
enhancement on the stellar tracks by means of the work of
\citet{Saletal00}.  Their stellar tracks are used in a differential
way in order to be independent of intrinsic differences with respect
to the stellar tracks of \citet{CCC97}. More specifically, we use the
fluxes in the lines and in the continua of the SSP models of
\citet{Maretal03} computed with the solar-scaled and the \aFe\
enhanced stellar tracks of \citet{Saletal00}, in order to determine
the fractional flux changes caused by the \aFe\ enhanced stellar
tracks.  Model~2 is then obtained by applying these fractional changes
to the fluxes in the lines and continua of Model~1.

\section{Comparison with elliptical galaxies}
\begin{figure*}
\begin{minipage}{\linewidth}
\centering\includegraphics[width=0.55\textwidth]{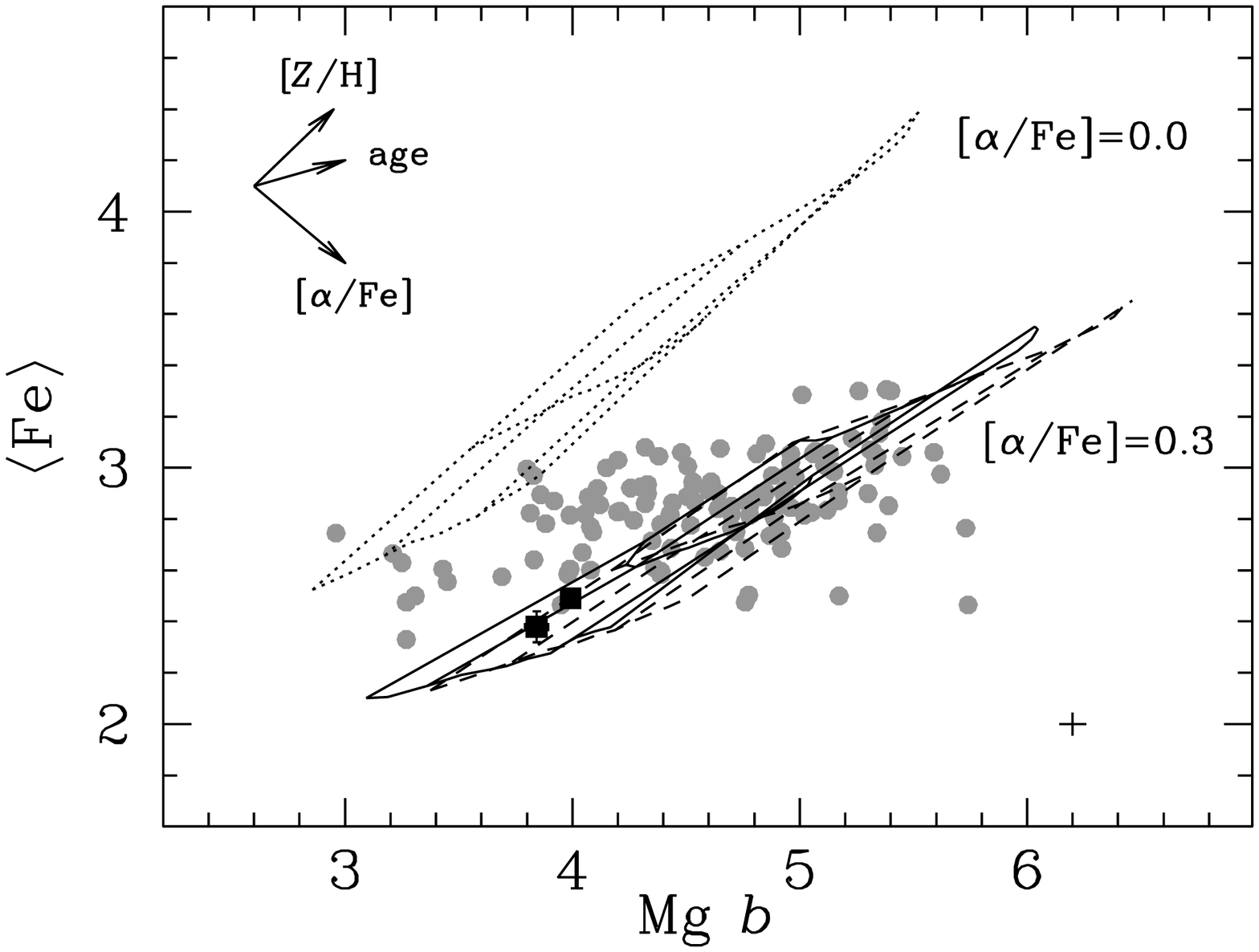}
\end{minipage}
\caption{\Mgb\ vs.\ \Fe\ for SSP models of constant ages
$t=3,5,10,15$~Gyr and metallicities $[\ZH]=0.0,0.35,0.67$. The dashed
and solid lines are the \aFe\ enhanced models Model~1 and Model~2, the
dotted lines are Model~3 (see Table~\ref{tab:models} and Sect.~3).
Grey circles are early-type galaxy data measured within $0.1\ r_{\rm
e}$ \citep{G93,Mehetal00,Beuetal02}. The typical error is given in the
lower-right corner. Black squares are the Bulge clusters NGC~6528 and
NGC~6553 \citep{Puzetal02}.}
\label{fig:enhanced1}
\end{figure*}
In this section we compare the Lick indices \Hb, \Mgb, and \Fe\ of the
three model flavors of Table~\ref{tab:models} with observational data
of early-type galaxies.

The data sample comprises 126 objects, 71 of which are field and 55
cluster objects, containing roughly equal fractions of elliptical and
lenticular (S0) galaxies. The sample is constructed from the following
sources: 41 Virgo cluster and field galaxies \citep{G93}, 32 Coma
cluster galaxies \citep{Mehetal00}, and 53 mostly field galaxies (the
highest quality data from \citealt{Beuetal02}) selected from the
ESO--LV catalog \citep{LV89}. For sake of homogeneity, in all cases
the central line-strength indices within 0.1~$r_{\rm e}$ are
considered.  A detailed assessment of the stellar population
parameters of the data and the discussion in the context of galaxy
formation is given in \citet{TMB02}.

\subsection{\aFe\ ratios}
Fig.~\ref{fig:enhanced1} shows Model~1 and Model~2 for various ages
and metallicities as dashed and solid lines in the \Mgb-\Fe\
plane. Grey circles are the early-type galaxy data quoted above.
Black squares are the two Bulge clusters NGC~6528 and NGC~6553 from
\citet{Puzetal02}.

The inclusion of \aFe\ enhanced stellar tracks (Model~2) has only a
negligible impact on the models in the \Mgb-\Fe\ plane. This is
consistent with the conclusion of \citet{Maretal03} for the
solar-scaled models that the decrease of metallic line strengths
like \Mgb\ and \Fe\ due to the \aFe\ enhancement of the stellar tracks
is comparable. This ensures that consistent \aFe\ ratios are derived
irrespective of whether we adopt Model~1 or Model~2.

For comparison, the solar-scaled model (Model~3) is shown as dotted
lines in Fig.~\ref{fig:enhanced1}. It becomes obvious that the major
effect of element abundance variations is the direct response of the
absorption lines, as accounted for in the TMB models (Model~1).

\subsection{Ages and metallicities}
\begin{figure*}
\begin{minipage}{\linewidth}
\centering\includegraphics[width=0.95\textwidth]{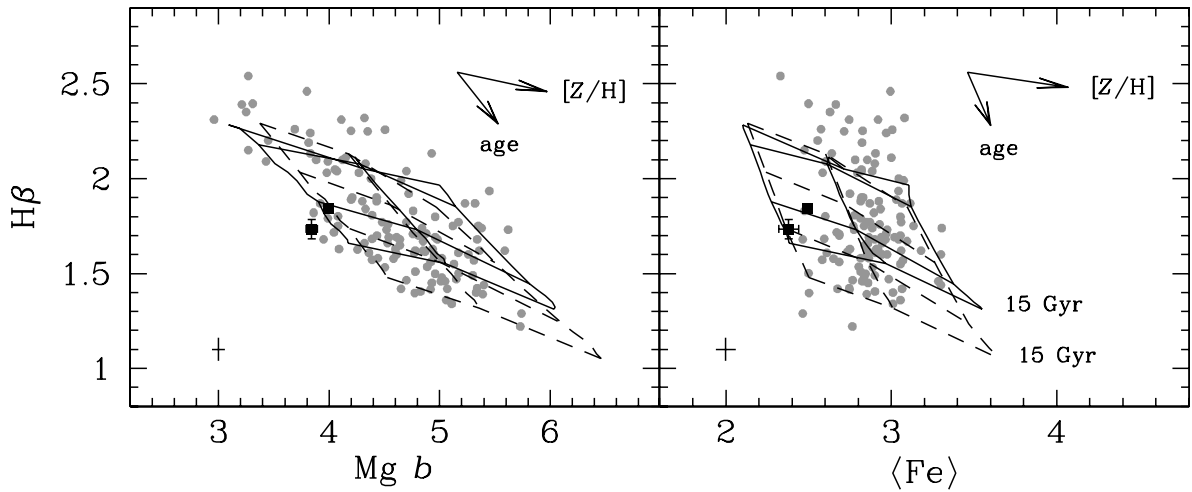}
\end{minipage}
\caption{\Mgb\ and \Fe\ vs.\ \Hb\ for SSP models of constant ages
$t=3,5,10,15$~Gyr and metallicities $[\ZH]=0.0,0.35,0.67$. The dashed
and solid lines are Model~1 and Model~2, respectively
(Table~\ref{tab:models}).  Grey circles are early-type galaxy data
measured within $0.1\ r_{\rm e}$ \citep{G93,Mehetal00,Beuetal02}. The
typical error is given in the lower-left corner. Black squares are the
Bulge clusters NGC~6528 and NGC~6553 \citep{Puzetal02}.}
\label{fig:enhanced2}
\end{figure*}
In Fig.~\ref{fig:enhanced2}, \Mgb\ and \Fe\ of Model~1 and
Model~2 are plotted vs.\ \Hb.  Consistent with \citet{Maretal03},
Model~2 has significantly stronger \Hb\ lines (by up to $\sim
0.26$~\AA\ at 15~Gyr) than Model~1, because of the hotter turnoffs of
the \aFe\ enhanced stellar tracks.  The strengthening of \Hb\
conspires with the weakening of the metallic lines such that going
from Model~1 to Model~2 the grid moves along the lines of constant
metallicity. Apparently, in current stellar evolution models enhancing
the \aFe\ element ratio and decreasing the age have very similar
effects on a stellar isochrone. This implies that the same
metallicities are derived from Model~1 and Model~2.

In contrast, ages are severely affected. Model~2 predicts
significantly stronger \Hb\ indices at the same age and metallicity,
which leads to the derivation of considerably older ages.  The
decision whether Model~1 or Model~2 is more appropriate relies on the
calibration with globular clusters for which ages and element
abundances are independently known.  The ages derived from Balmer
lines for the two solar metallicity Bulge globular clusters NGC~6528
and NGC~6553 with SSP models using \aFe\ enhanced stellar tracks are
in better agreement with the ages inferred from color-magnitude
diagrams \citep{Maretal03}. This is visualized in
Fig.~\ref{fig:enhanced2} with the globular clusters being the filled
squares. However, the uncertainties are too large to allow for a firm
conclusion in favor of Model~2.

Since the relative increase of the \Hb\ line-strength due to \aFe\
enhanced tracks increases with metallicity, the discrepancy between
the ages obtained with Model~1 and Model~2 is more pronounced for
elliptical galaxies than for the most metal-rich globular clusters.  A
considerable fraction of the data lies significantly below the 15~Gyr
line of Model~2 (Fig.~\ref{fig:enhanced2}).  Model~2 yields ages up to
$\sim 30$~Gyr (in extrapolation), while with Model~1 we obtain ages
not older than $\sim 15$~Gyr for the same data set.

%========================= S U M M A R Y =================================

\section{Discussion and Conclusion}
We complement our study of \aFe\ enhanced stellar population models of
Lick absorption indices \citep[][TMB]{TMB03} by comparing two sets of
\aFe\ enhanced models. In both models the major effect from \aFe\
enhancement on Lick indices is accounted for through a modification of
the stellar absorption line-strengths using the response functions of
\citet{TB95}.  One set of models, however, uses solar-scaled (Model~1,
equivalent to the TMB model), the other \aFe\ enhanced (Model~2)
stellar evolutionary tracks. Here we compare the \aFe\ ratios,
metallicities and ages that are derived for elliptical galaxies from
Model~1 and Model~2.

Because of the hotter red giant branches of the \aFe\ enhanced stellar
tracks, Model~2 has weaker metallic index line-strengths than Model~1.
The hotter turnoffs of the \aFe\ enhanced stellar tracks lead to
considerably larger Balmer line indices (\Hb) in Model~2. Both \Mgb\
and \Fe\ indices are reduced almost equally, so that the determination
of \aFe\ ratios for stellar systems is not significantly affected. The
decrease of the metallic indices and the increase of \Hb\ conspire in
a way that also consistent metallicities are obtained with Model~1 and
Model~2.

The age derivation for metal-rich stellar populations, instead, is
severely affected by the use of \aFe\ enhanced stellar tracks. The
stronger \Hb\ lines of Model~2 with respect to Model~1 imply
significantly older ages.  The higher the metallicity the more
pronounced is this increase of \Hb\ in Model~2, and average ages up to
$\sim 30$~Gyr are obtained for early-type galaxies.

Such high ages appear larger than current estimates of the age of the
universe \citep*[$\sim 15$~Gyr,][]{FMS01} as constrained by cosmic
microwave background \citep{KCS01} and high-redshift supernova
\citep{Rieetal98,Peretal99} measurements. This may indicate that the
bluing of the stellar tracks due to \aFe\ enhanced chemical mixtures
at super-solar metallicities is overestimated in current stellar
evolution models. 

The key check, whether \aFe\ enhancement is described adequately in
the stellar evolutionary tracks, is the comparison with globular
cluster data. However, at metallicities well below solar, where plenty
of data of globular clusters are available, the effect of \aFe\
enhancement on stellar evolution is negligible. Around solar
metallicity the effect is appreciable, but the presently available
calibrations do not provide a clear answer. The comparison with the
color-magnitude-diagram of 47 Tuc suggests that the Red Giant Branch
of the \citet{Saletal00} \aFe\ enhanced tracks is indeed too warm
\citep{Schetal02}.  The integrated absorption indices of the two Bulge
clusters NGC~6528 and NGC~6553 are consistent with both \aFe\ enhanced
and solar-scaled stellar tracks, as discussed before.  At
metallicities above solar, where the impact of \aFe\ enhanced stellar
tracks becomes most significant and the ages of ellipticals become
alarmingly high, globular clusters are not available.  We are trapped
in the usual dilemma for the stellar population models of elliptical
galaxies.  Hence, further investigations on the effect of \aFe\
enhanced mixtures on stellar evolutionary tracks at super-solar
metallicities would be very valuable.

%========================= R E F E R E N C E S ===========================

%\bibliography{mnrasmnemonic,literature}
\normalsize

\end{document}